\pdfoutput=1
\documentclass[aps,prl,twocolumn, amsfonts,showpacs, superscriptaddress]{revtex4}
\usepackage{mathptmx}
\usepackage{epsfig,amsopn}
\usepackage{graphicx}
\usepackage{amsmath,amssymb}
\usepackage{natbib}
\usepackage{braket}

\newcommand{\eref}[1]{Eq.~(\ref{#1})}%
\newcommand{\fref}[1]{Fig.~\ref{#1}} %
\newcommand{\Fref}[1]{Figure~\ref{#1}}%
\makeatletter

\renewenvironment{widetext@grid}{%
  \par\ignorespaces
  \setbox\widetext@top\vbox{%
   \vskip15\p@
   \hb@xt@\hsize{%
    \leaders\hrule\hfil
    \vrule\@height6\p@
   }%
   \vskip6\p@
  }%
  \setbox\widetext@bot\hb@xt@\hsize{%
    \vrule\@depth6\p@
    \leaders\hrule\hfil
  }%
  \onecolumngrid
  \let\set@footnotewidth\set@footnotewidth@ii
}{%
  \par
  \twocolumngrid\global\@ignoretrue
  \@endpetrue
}%

\makeatother

\begin{document}

\title{Universal Large Deviations for the Tagged Particle in  Single
  File Motion} 

\author{Chaitra Hegde} \author{Sanjib Sabhapandit}  
\affiliation{Raman Research Institute,  Bangalore - 560080, India}
\author{Abhishek Dhar}
\affiliation{International centre for theoretical sciences, TIFR,
Bangalore - 560012, India}

\date{\today}

\begin{abstract}
We consider a gas of point particles moving in a one-dimensional channel with a
hard-core inter-particle interaction that prevents particle crossings
--- this is called single-file motion. Starting from equilibrium initial
conditions we observe the motion of a tagged particle. It is well
known that if the individual particle dynamics is diffusive, then the
tagged particle motion is sub-diffusive, while for ballistic particle
dynamics, the tagged particle motion is diffusive. Here we compute
exactly the large deviation function for the tagged particle
displacement and show that this is universal, independent of the
individual dynamics.
\end{abstract}
\maketitle

\begin{figure}
\includegraphics[width=3.2in]{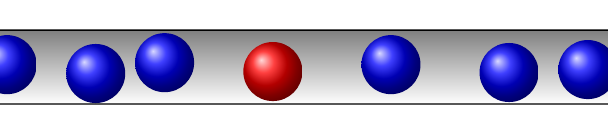} 
\caption{(Color online) A schematic diagram of single-file motion of
  particles in a narrow channel where they cannot pass each other. We
  study the motion of a single tagged particle (say the red colored
  one).}
\label{schematic}
\end{figure}

The motion of particles in narrow channels where the particles cannot
overtake each other is referred to as single-file motion [see
Fig.~(\ref{schematic})]. This concept was introduced by Hodgkin and
Keynes~\cite{hodgkin55} to describe ion transport in biological
channels.  The motion of a tagged particle in such a single-file
system has been of great interest since the classic papers by
Jepsen~\cite{jepsen65} and Harris~\cite{harris65}. These papers showed
that, in a gas of hard rods evolving with Hamiltonian dynamics, a
tagged particle moves diffusively~\cite{jepsen65} with the mean square
displacement (MSD) growing linearly with time $t$, whereas for a gas
of Brownian particles, the tagged particle shows
sub-diffusion~\cite{harris65} with the MSD growing as $ \sqrt{t}$.
There has been a revival of interest in tagged particle diffusion as
several experiments are now able to observe this in single-file
systems in both colloidal and atomic single-file
systems~\cite{hahn96,kukla96,wei00,lutz04,lin05,das10}, and some of
the theoretical predictions have been verified.

There have been a number of studies to understand tagged particle
motion in systems with deterministic as well as stochastic
dynamics~\cite{lebowitz67,lebowitz72,
percus74,beijeren83,arratia83,pincus78,rodenbeck98,
majumdar91,lizana08,barkai09,
kollmann2003,gupta07,barkai10,roy13,roy14,
sabhapandit:07,illien13,benichou13}.  Attempts have been made to
obtain the full probability density function (PDF) for the tagged
particle displacement. The $N$-particle propagator has been obtained
using the ``reflection principle''~\cite{rodenbeck98} and Bethe
Ansatz~\cite{lizana08}, and from this the tagged particle distribution
has been obtained by integrating out all other particles. However, the
resulting form of the distribution is complicated and not very
illuminating.  An approximate scheme relying on Jepsen's mapping to
non-interacting particles has been used in~\cite{barkai09,barkai10}.
A recent work~\cite{Krapivsky:14} has used macroscopic fluctuation
theory~\cite{lasinio} to compute the cumulant generating function
(CGF) corresponding to the tagged particle PDF.

In this Letter, we show that it is possible to exactly compute the
large time asymptotic form of the PDF of tagged particle displacement.
Our method is applicable to deterministic as well as stochastic
systems that are initially in equilibrium. This leads to a universal
form for the PDF.  We consider a collection of hard-point identical
particles distributed with an uniform density $\rho$ on the one
dimensional line from $-\infty$ to $\infty$. Each particle moves
independently using the same dynamics, except that the hard-core
repulsion prevents crossing of particles.  We consider a
single-particle propagator of the general form
\begin{equation}
G(y,t|x,0)=
\frac{1}{\sigma_t}\,f\left(\frac{y-x}{\sigma_t}\right),
\label{propagator} 
\end{equation}
where $f(-w)=f(w) \ge 0$, $\int_{-\infty}^\infty f(w)\, dw=1$, and
$\langle |y-x|\rangle/\sigma_t=\int_{-\infty}^\infty |w| f(w)\,
dw=\Delta$ is finite.  Using a mapping to the non-interacting gas
picture, we show that the PDF of the displacement $X_t$, of the tagged
particle, has the large deviation form
\begin{equation}
P_\mathrm{tag}(X_t,t|0,0) \sim e^{-\rho \sigma_t I(X_t/\sigma_t)},
\label{LDform}
\end{equation}
where the large deviation function (LDF) is given exactly by
\begin{subequations}
\begin{align}
\label{I(z)}
I(z)&=2 Q(z)-\bigl[4 Q^2(z)-z^2\bigr]^{1/2},\\
\label{Q(z)}
\text{with}\quad Q(z)&= z\int_0^z f(w)\, dw + \int_z^\infty w f(w)\, dw.
\end{align}
\end{subequations}
We also compute the leading order correction exactly [see
Eqs.~\eqref{analytic} and \eqref{gz}].

We first outline the strategy used in the calculation.  Initially, we
consider $2N+1$ particles, independently and uniformly distributed in
the interval $[-L,L]$.  In the computation, we assume both $N$ and $L$
to be large and keep only the dominant term. Finally, we take the
limit $N\to\infty$, $L\to\infty$ while keeping $N/L=\rho$ fixed.
Since during a collision each particle acts as a reflecting hard wall
for the other and the particles are identical, one can effectively
treat the system of the interacting hard-point particles as
non-interacting by exchanging the identities of the particles emerging
from collisions. In the non-interacting picture, each particle
executes an independent motion and the particles \emph{pass through
each other} when they `collide'.  The position of each particle at
time $t$ is given independently by the propagator
in \eref{propagator}.  In many physical problems, the propagator is
Gaussian, i.e., $f(x)=e^{-x^2/2}/\sqrt{2\pi}$, where $\sigma_t$ is
simply the standard deviation.  For example, for Hamiltonian dynamics
with initial velocities chosen independently from Gaussian
distribution with zero mean and variance $\bar{v}^2$ we have
$\sigma_t=\bar{v} t$.  On the other hand, for Brownian particles,
$\sigma_t=\sqrt{2 D t}$, where $D$ is the diffusion coefficient. For
fractional Brownian motion, $\sigma_t\propto t^H$, where $H$ is the
Hurst exponent. However, our analysis is valid for a general
propagator. Note that the dependence on time only appears through the
characteristic displacement $\sigma_t$ in time $t$.

The joint probability density of the middle tagged particle being
at $x$ at time $t=0$, and at $y$ at time $t$, can be expressed in terms
of properties of the non-interacting particles. In the non-interacting
picture, there are two possibilities: (i) the middle particle at time
$0$ is still the middle particle at time $t$, (ii) a second particle
has become the middle particle at time $t$.  We need to sum over these
two processes.

To compute the contribution from process (i) we pick one of the
non-interacting particles at random with a density $\rho$, multiply by
the propagator [\eref{propagator}] that it goes from $(x, 0)$ to
$(y,t)$, and then multiply by the probability that it is the middle
particle at both $t=0$ and $t$.  Thus one obtains:
\begin{equation}
P_{(1)}(x,0;y,t)= 
\rho \,G(y,t|x,0)\,F_{1N}(x, y, t),
\label{jep1-1}
\end{equation} 
where $F_{1N}(x, y, t)$ is the probability that there are an equal
number of particles to the left and right of $x$ and $y$ at $t=0$ and $t$
respectively.

To compute the contribution from process (ii), we first pick two
particles at random at time $t=0$, and multiply by the propagators
that they go from $(x,0)$ to $(\tilde y,t)$ and $(\tilde x, 0)$ to
$(y, t)$ respectively. We then multiply by the probability there are
an equal number of particles on both sides of $x$ and $ y$ at $t=0$
and $t$ respectively. Finally, integrating with respect to
$\tilde{x},\tilde{y}$, we get
\begin{align}
P_{(2)}(x,0;y,t)= &
\rho^2
\int_{-\infty}^\infty d\tilde x \int_{-\infty}^\infty 
 d\tilde y \notag \\  \times &G(\tilde y,t|x,0)\,G(y,t|\tilde{x},0) \,F_{2N}(x,y, \tilde
 x,\tilde y, t),   
\label{p2eq}
\end{align}
where $F_{2N}(x,y, \tilde x,\tilde y, t)$ is the probability that
there are an equal number of particles on both sides of $x$ and
$ y$ at $t=0$ and $t$ respectively, given that there is a  
particle at $\tilde x$ at time $t=0$, and a particle at $\tilde y$ at time $t$.
The  joint PDF of the tagged particle  is exactly given by 
\begin{equation}
P(x,0;y,t)= P_{(1)}(x,0;y,t)+P_{(2)}(x,0;y,t)~. \label{ptot}
\end{equation}

To proceed further, we need the expressions for $F_{1N}$ and $F_{2N}$.
Let $p_{-+}(x, y,t)$ be the probability that a particle is to the left
of $x$ at $t=0$ and to the right of $y$ at time $t$.  Similarly, we
define the other three complementary probabilities. Clearly,
\begin{subequations}
\begin{align} 
p_{-+}(x,y,t)&=(2L)^{-1}\int_{-L}^x dx' \int_y^{\infty} dy'
G(y',t|x',0),\\
p_{+-}(x,y,t)&=(2L)^{-1}\int_x^{L} dx' \int_{-\infty}^y dy' G(y',t|x',0),\\
p_{--}(x,y,t)&=(2L)^{-1}\int_{-L}^x dx' \int_{-\infty}^y dy' G(y',t|x',0),\\
p_{++}(x,y,t)&=(2L)^{-1}\int_x^{L} dx' \int_y^{\infty} dy' G(y',t|x',0),
\label{transp} 
\end{align} 
\end{subequations}
and $p_{++}+p_{+-} + p_{-+} + p_{--} = 1$. In terms of these
probabilities, $F_{1N}$ can be expressed as~\cite{suppl},
\begin{equation*}
F_{1N}(x,y, t) = 
\int_{-\pi}^{\pi} \frac{d\phi}{2\pi}\int_{-\pi}^{\pi}\frac{d\theta}{2\pi} 
\,\bigl[H(x,y,\theta,\phi,t)\bigr]^{2N} ,
\end{equation*}
where
\begin{align}
H(x,y,\theta,\phi,t)&=p_{++}(x,y, t) e^{i\phi} + p_{--}(x,y, t)
e^{-i\phi} \notag\\
&+ p_{+-}(x, y, t) e^{i\theta} + p_{-+}(x, y, t) e^{-i\theta}.
\end{align}
The angular integrals enforce the condition that the total number of
particles crossing the middle particle from left-to-right is the same
as the total number from right-to-left. This can be seen by explicitly
performing the multinomial expansion above and computing the angular
integrals. Using the fact that $2N$ is even and the integrand is
unchanged if both $\theta$ and $\phi$ are shifted by $\pi$ we can
write $F_{1N}$ in the form
\begin{equation}
F_{1N}(x,y, t) = 
\int_{-\pi/2}^{\pi/2} \frac{d\phi}{\pi}\int_{-\pi}^{\pi}\frac{d\theta}{2\pi} 
\,\bigl[H(x,y,\theta,\phi,t)\bigr]^{2N} .
\label{jep1-3}
\end{equation}

Similar argument can be used to compute $F_{2N}$. However, in this
case, one has to keep track of the order of the positions
$(x,\tilde{x})$ and $(y,\tilde{y})$. One finds~\cite{suppl}
\begin{align}
F_{2N}(x,y,\tilde{x},\tilde{y},t) = 
\int_{-\pi/2}^{\pi/2} \frac{d\phi}{\pi}\int_{-\pi}^{\pi} &\frac{d\theta}{2\pi} 
\,\bigl[H(x,y,\theta,\phi,t)\bigr]^{2N-1} \notag \\ &\times \psi(\theta,\phi|x,y,\tilde{x},\tilde{y}) ,
\label{jep1-3}
\end{align}
where the extra phase factor is given piece-wise by
$\psi=e^{-i\phi}$, $e^{i\phi}$, $e^{-i\theta}$,  and
$e^{i\theta}$ for the situations
 (a) $\tilde x <x$ and $\tilde{y} < y$,
 (b) $\tilde x >x$ and $\tilde{y} > y$,
 (c) $\tilde x <x$ and $\tilde{y} > y$, and
 (d) $\tilde x >x$ and $\tilde{y} < y$ respectively.

Now, substituting the above form of $F_{2N}$ in \eref{p2eq}, and
performing the integration over $\tilde x$ and $\tilde y$,  while using the
property $G(y,t|x,0)=G(y-x,t|0,0)$, we get
\begin{align}
&P_{(2)}(x,0;y,t)= 
\rho^2 \int_{-\pi/2}^{\pi/2} \frac{d\phi}{\pi}\int_{-\pi}^{\pi}\frac{d\theta}{2\pi} 
\,\bigl[H(x,y,\theta,\phi,t)\bigr]^{2N-1}\notag\\
&\qquad\times\bigl[2 A_1(z) A_2(z) \cos\phi + A_1^2(z) e^{-i\theta} +
  A_2^2(z)e^{i\theta}\bigr] ,
\end{align}
where $z=(y-x)/\sigma_t$ and the functions $A_{1,2}(z)$ are given by
\begin{subequations}
\begin{align}
A_1(z)&=\int_{\sigma_t z}^\infty G(x,t|0,0)\, dx=\int_z^\infty f(w)\, dw,\\
\text{and}~~A_2(z)&=1-A_1(z).
\end{align}
\end{subequations}

Now we explicitly compute the expressions for $p_{\pm\pm}$ using
\eref{propagator}. Keeping only the dominant terms up to  $O(1/L)$,
which survive in the limit $N\to \infty,~L \to \infty$ while keeping
$N/L=\rho$ fixed, we get
\begin{subequations}
\begin{align}
p_{-+}&=\frac{\sigma_t}{2L}\left[-\frac{z}{2} + Q(z) \right] +\dotsb\\
  p_{+-}&=\frac{\sigma_t}{2L}\left[\frac{z}{2} + Q(z) \right]
  +\dotsb\\ p_{--}&=\frac{1}{2}
  +\frac{\sigma_t}{2L}\left[\frac{\bar{z}}{2} - Q(z) \right]
  +\dotsb \\ p_{++}&=\frac{1}{2}
  +\frac{\sigma_t}{2L}\left[-\frac{\bar{z}}{2} - Q(z) \right] +\dotsb,
\end{align}
\end{subequations}
where  $z=(y- x)/\sigma_t$, $\bar{z}=(y+ x)/\sigma_t$, and 
the function $Q(z)$ is given by \eref{Q(z)}.

To compute $H^{2N}$ for large $N$, it is useful to express $H$ in
the form
\begin{align}
H&=1-(1-\cos{\phi}) ~(p_{++}+ p_{--})+ i \sin \phi ~(p_{++}-p_{--})
\notag \\ 
&-(1-\cos {\theta}) 
(p_{+-} + p_{-+} ) +i \sin \theta~(p_{+-}-p_{-+}).
\end{align} 
Now, substituting $p_{\pm\pm}$ in the above expression of $H$, for
large $N$, keeping only the most dominant terms, one finds
\begin{equation}
H^{2N}= e^{-2N (1-\cos\phi)}
e^{-i  \rho \sigma_t \bar{z} \sin \phi}
e^{-2 \rho\sigma_t Q(z) (1-\cos \theta)} e^{i \rho\sigma_t z \sin \theta}.
\end{equation}

Thus we have explicitly obtained $P_{(1)},P_{(2)}$ and hence
$P(x,0;y,t)$ defined in \eref{ptot}.  Using this we can finally write
down the propagator for the displacement $X_t=y-x$ of the tagged
particle as
\begin{math}
P_\text{tag}(X_t,t|0,0)=\int\int \delta\bigl(X_t
  -[y-x] \bigr)\, P(x,0;y,t)\, dx\, dy.
\end{math}
Now making a change of variables from $x,y$ to $z, \bar{z}$, we get 
\begin{align}
&P_\text{tag}(X_t=\sigma_tz,t|0,0)=\notag\\
&\lim_{N\to\infty} \int_{-\infty}^\infty \frac{d\bar{z}}{2}
\int_{-\pi/2}^{\pi/2}
\frac{d\phi}{\pi}\int_{-\pi}^{\pi}\frac{d\theta}{2\pi} 
\rho B(z,\theta,\phi)\notag\\
&\times e^{-2N (1-\cos\phi)}
e^{-i  \rho \sigma_t \bar{z} \sin \phi}
e^{-2 \rho\sigma_t Q(z) (1-\cos \theta)} e^{i \rho\sigma_t z \sin \theta},
\end{align} 
where 
\begin{math}
B(z,\theta,\phi)=f(z) +\rho\sigma_t
\bigl[2 A_1(z) A_2(z) \cos\phi + A_1^2(z) e^{-i\theta} +
  A_2^2(z)e^{i\theta}\bigr] .
\end{math}
For large $N$, the major contribution of the integral over $\phi$
comes from the region around $\phi=0$. Therefore, the $\phi$ integral
can be performed by expanding around $\phi=0$ to make it a Gaussian
integral (while extending the limits to $\pm\infty$). Subsequently,
one can also perform the Gaussian integral over $\bar{z}$. This leads
to the exact expression, 
\begin{align}
P_\text{tag}(X_t=\sigma_t z,t|0,0)=
\frac{1}{\sigma_t}
\int_{-\pi}^{\pi}\frac{d\theta}{2\pi} \,
B(z,\theta)\notag \\
\times e^{-\rho\sigma_t \bigl[2  Q(z) (1-\cos \theta)-i  z \sin \theta\bigr]},
\label{P_exact}
\end{align}
where
\begin{align}
&B(z,\theta)\equiv B(z,\theta,0)\notag \\
& = f(z)+
\rho\sigma_t
\bigl[2 A_1(z) A_2(z) + A_1^2(z) e^{-i\theta} +
  A_2^2(z)e^{i\theta}\bigr].
\end{align}
Since $\sigma_t$ is an increasing function of time, the integral over
$\theta$ can be evaluated for large $t$, using the saddle point
approximation. This gives the large deviation form given
by \eref{LDform} with the large deviation function given by
\begin{equation*}
I(z)= 2 Q(z)(1-\cos \theta^*) -i z \sin
\theta^*,
~\text{with}~\tan\theta^*=\frac{i z}{2Q(z)}.
\end{equation*}
Eliminating $\theta^*$  yields the  form given by \eref{I(z)}. 
The full asymptotic form of the propagator of the tagged particle
displacement, obtained from the saddle point approximation is
\begin{equation}
P_\text{tag}(X_t=\sigma_t z,t|0,0) \approx
\frac{1}{\sigma_t}
\frac{\sqrt{\rho\sigma_t}}{\sqrt{2\pi g_2(z)}}\,
g_1(z)\,e^{-\rho\sigma_t I(z)} ,
\label{analytic}
\end{equation}
where $g_2(z)=\bigl[4 Q^2(z)-z^2\bigr]^{1/2}$ has come from performing
     the Gaussian integral around the saddle-point $\theta^*$ and
\begin{align}
&g_1(z)=  (\rho\sigma_t)^{-1} \, B(z,\theta^*)\notag\\
&=2 A_1(z) A_2(z)  + A_1^2(z) \frac{\sqrt{2Q(z)+z}}{\sqrt{2Q(z)-z}} +
  A_2^2(z) \frac{\sqrt{2Q(z)-z}}{\sqrt{2Q(z)+z}} \notag\\
&\quad+ O\bigl([\rho\sigma_t]^{-1}\bigr). \label{gz}
\end{align}
Note that the process (i) where, in the non-interacting picture, the
same particle happens to be the middle particle at both the initial
and final times, does not contribute at this order, but to
$O([\rho \sigma_t]^{-1})$ in the expression of $g_1(z)$.  In fact, one
can systematically obtain the corrections to the above expression of
$g_1(z)$, order by order.  By keeping terms beyond the second-order in
the expansion of the argument of the exponential function around the
saddle-point $\theta^*$ in \eref{P_exact}, subsequently expanding the
exponentials of the higher order terms in power series, and also
expanding $B(z,\theta)$ around $\theta^*$ in power series, the
resulting integrals in \eref{P_exact} are exactly doable in terms of
gamma functions.  In the limit $z\to 0$ we get $g_1(0)=1$,
$g_2(0)=2Q(0)=\Delta$ and $I(z)=z^2/(2\Delta) + O(z^4)$. Therefore, in
this limit,
\eref{analytic} reduces to a Gaussian form with a variance
\begin{equation}
\langle X_t^2 \rangle_c =\frac{\Delta\sigma_t}{\rho},
\label{Percus} 
\end{equation}
which is the so-called Percus relation~\cite{percus74, barkai10}.  The
corrections to this result can obtained following a similar proceedure
explained above [between Eqs.~\eqref{gz} and \eqref{Percus}].  The
Gaussian form is expected to hold near the central region $|X_t|
\lesssim O(\sqrt{\sigma_t/\rho})$. However, away from this central
region, the Gaussian approximation breaks down and one needs the
complete form given by \eref{analytic}.  

\begin{figure}
\includegraphics[width=3.2in]{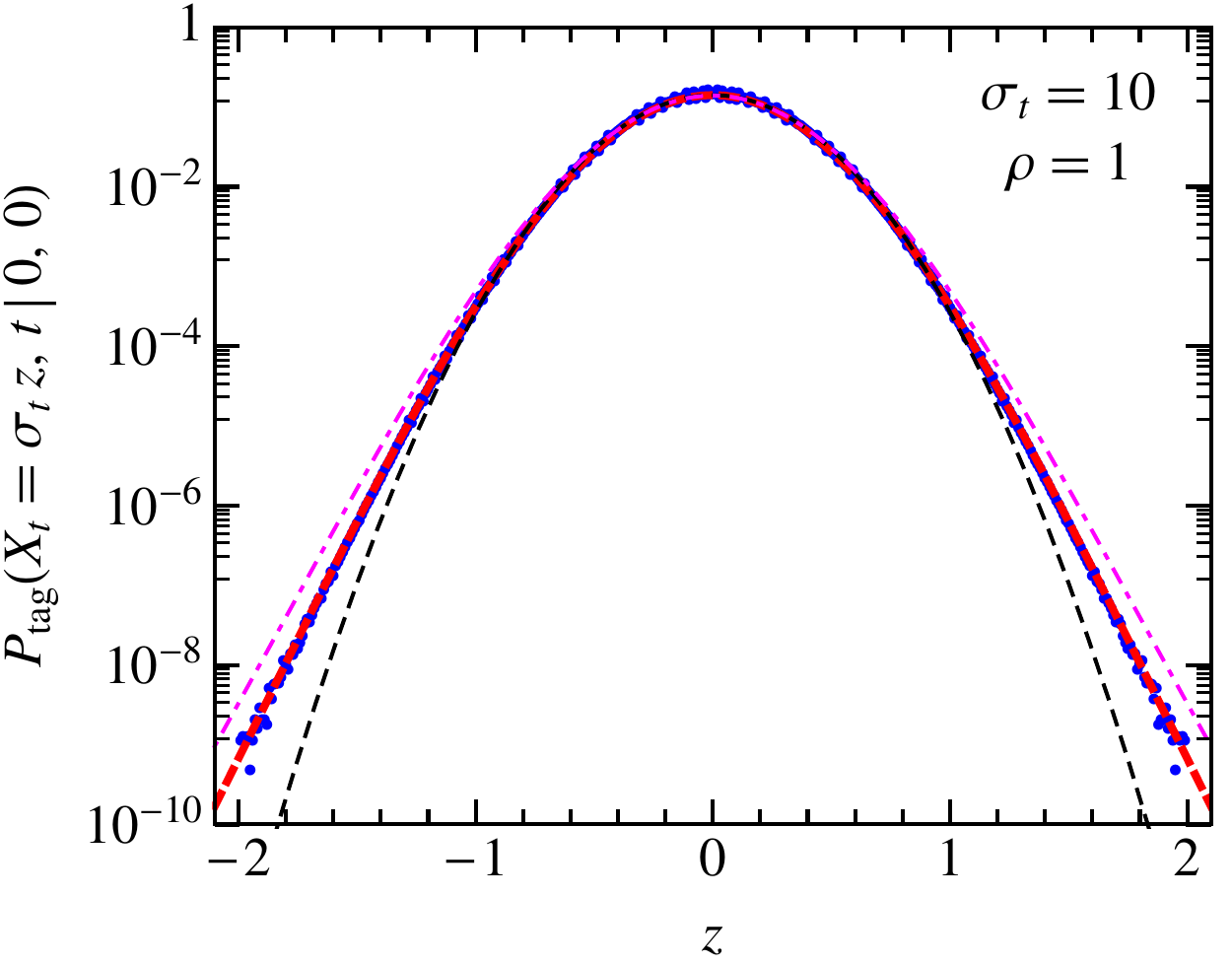} 
\caption{(Color online) The (blue) points represent the simulation
results for the PDF of the displacement of the tagged middle particle
in a gas of $2N+1$ particles initially distributed uniformly in a box
between $[-N,N]$ with $N=1000$.  Using the mapping to the
non-interacting picture, each particle is evolved independently
according to the Gaussian propagator with $\sigma_t=10$ and the
difference in positions of the middle particles at the initial and the
final times respectively is the displacement of the tagged middle
particle in the interacting particle system.  The PDF is computed
using $32\times 10^9$ realizations.  The (red) thick dashed line
corresponds to the analytic result in Eq.~(\ref{analytic}), while the
(magenta) dot-dashed line plots the (numerically normalized) large
deviation form given by \eref{LDform}. The (black) dashed line is
Gaussian distribution with the variance given by \eref{c2}.}
\label{pdf}
\end{figure}

For a Gaussian propagator, we explicitly get
\begin{equation*}
Q(z)=\frac{e^{-z^2/2}}{\sqrt{2\pi}}  +\frac{z}{2}\,
\mathrm{erf}\bigl(z/\sqrt{2}\bigr), 
~\text{and}~A_1(z)=\frac{1}{2}\mathrm{erfc}\bigl(z/\sqrt{2}\bigr).
\end{equation*}
Using these expressions, in \fref{pdf}, we plot the (numerically
normalized) large deviation form given by \eref{LDform}, the complete
form given by \eref{analytic} and its Gaussian approximation, and
compare them with numerical simulation results.  We note that the
large deviation form of the PDF, given by \eref{LDform}, really
implies the mathematical equality
\begin{equation*}
I(x)=-\lim_{\sigma_t \to \infty} \frac{1}{\rho\sigma_t}\ln P_\mathrm{tag}(X_t=x\sigma_t)~.
\end{equation*}
However, to achieve the required large time limit for comparison with
real data is often difficult, and it is necessary to include the
sub-leading correction.  Indeed, \eref{analytic}, which includes the
correction term, agrees extremely well with the numerical simulation
results.  We note that, for diffusive systems, our result can be
recovered by taking appropriate limits of the corresponding
expressions in \cite{rodenbeck98}.

Now, we look at the cumulant generating function of the tagged
particle displacement $X_t$, defined through
\begin{equation}
Z(\lambda)
=\Bigl\langle e^{\lambda \rho X_t}\Bigr\rangle
=e^{\rho \sigma_t \mu(\lambda)}. 
\end{equation}
Using the large deviation form of $P_\mathrm{tag}(X_t,t|0,0)$ given
by \eref{LDform}, and then evaluating the integral over $z$ using the
saddle point approximation, we have $\mu(\lambda)=\lambda z^* -
I(z^*)$ where $z^*$ is implicitly given by the equation
$\lambda=I'(z^*)$. Using the expression of $I(z)$ obtained above in
terms of $\theta^*$ with the substitution $\theta^*=iB$, we can
express $\mu(\lambda)$ in the parametric form
\begin{subequations}
\begin{align}
\label{mu}
&\mu(\lambda)=\left[\lambda+\frac{1-e^B}{1+e^B} \right] z, \\
\label{lambda}
&\lambda=\Bigl(1-e^{-B}\Bigr)
\left[1+ \Bigl(e^B-1\Bigr)\,A_1(z) \right], \\
\label{B}
&e^{2B}=\frac{2Q(z)+z}{2Q(z)-z}.
\end{align}
\end{subequations}

\begin{figure}
\includegraphics[width=3.2in]{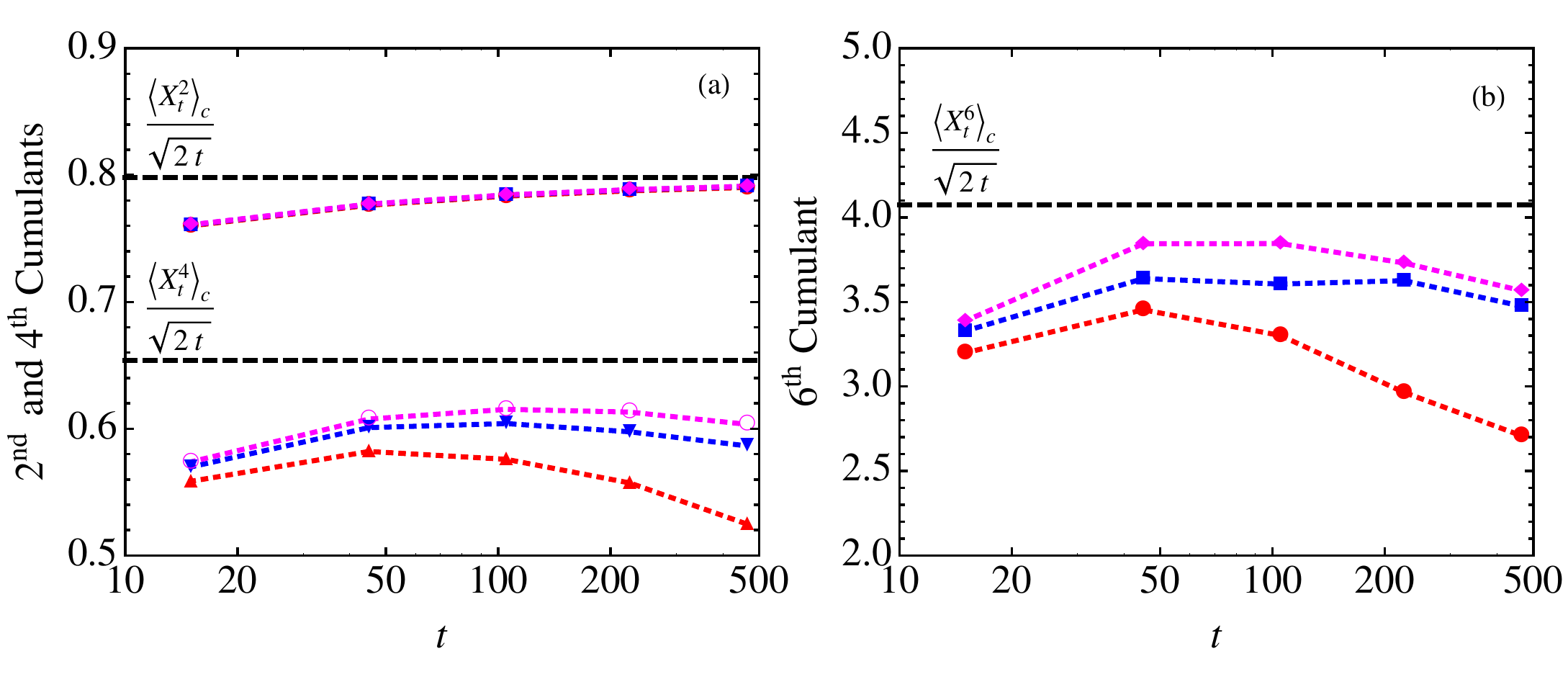} 
\caption{(Color online) Points connected by dotted lines are the simulation results for (a) $2^{\rm nd}$, $4^{\rm th}$
  and (b) $6^{\rm th}$ cumulants (scaled) of the displacement of the
  tagged middle particle in a gas of $2N+1$ hard point diffusing
  particles ($D=1$), initially distributed uniformly in a box between
  $[-N,N]$. The data is for system sizes $N=250$ (red, lowest curve),
  $500$ (blue, middle curve) and $750$ (magenta, upper curve). All
  cumulants are seen to approach our theoretical predictions (black
  dashed lines) with increasing system size.}
\label{moms}
\end{figure}

For the case of the Gaussian propagator with a variance $\sigma_t^2$,
 the first three even cumulants can be obtained as
\begin{subequations}
\begin{align}
\label{c2}
\langle X_t^2 \rangle_c &=\frac{\sqrt 2}{\rho\sqrt\pi}\sigma_t, \\
\label{c4}
\langle X_t^4 \rangle_c &= \frac{3 \sqrt{2}
  (4-\pi)}{(\rho\sqrt\pi)^{3}} \sigma_t~, \\
\label{c6}
\langle X_t^6 \rangle_c &=
\frac{15 \sqrt{2} \bigl(68-30\pi+3\pi^2\bigr)}{(\rho\sqrt\pi)^{5}} \sigma_t.
\end{align}
\end{subequations}
\Fref{moms} compares the above analytic expressions
 with the simulation results, for the case where individual particle
motion is diffusive.  Note that at large times, finite size effects
kick in and the curves start deviating from the expected infinite size
behavior. The higher cumulants sense the boundary effects at earlier
times than the lower ones.

In conclusion, we have explicitly computed exactly the large time
asymptotic form of the probability distribution of a tagged particle
in a single-file system and shown that this is universal.  This
unifies the treatment of single-file motion of particles with hardcore
interactions, within a general framework, as has also been attempted
in some earlier work~\cite{kollmann2003,barkai10}.  For the case of
Brownian particles, our results have been verified using macroscopic
fluctuation theory~\cite{Krapivsky:14}. However our microscopic
approach is more intuitive from a physical point of view, it is more
general, and directly gives the large deviation function as well as
the important corrections often required for comparison with real
data. The methods of the Letter can be extended to more general
initial conditions.

We thank K. Mallick, T. Sadhu, B. Derrida, and A. Roy for useful
discussions.  SS and AD acknowledge the hospitality of the GGI,
Florence during the workshop ``Advances in Nonequilbrium Statistical
Mechanics (2014)'' where part of this work was carried out. SS
acknowledges the support of the Indo-French Centre for the Promotion
of Advanced Research (IFCPAR/CEFIPRA) under Project 4604-3. AD thanks
DST for support through the Swarnajayanti fellowship.

\widetext
\bigskip
\noindent\rule{\hsize}{1.5pt}
\bigskip

\setlength{\baselineskip}{20pt}

\centerline{\large\bfseries Supplementary material for ``Universal
Large Deviations for the Tagged Particle in Single File Motion''}

\section{Evaluation of $F_{1N}(x,y,t)$}

Let $F_{1N}(x,y,t)$ be the probability that there are an equal number
of particles to the left and right of $x$ and $y$ at $t=0$ and $t$
respectively. In this case, one selected particle out of $2N+1$
particles, goes from $x$ to $y$ in time $t$. The remaining $2N$
particles are independent of each other and the selected particle. Let
$p_{-+}(x, y, t)$ be the probability that one of these particles is to
the left of $x$ at $t=0$ and to the right of $y$ at time $t$. Let
$p_{+-}$, $p_{--}$ and $p_{++}$ be similarly defined.

Let $n_1$ be the number of particles that go from the left of $x$ to the
left of $y$, $n_2$ be the number of particles going from  left to the right, $n_3$ be the number of particles going from the  right to the left, and $n_4$ be the number of particles going from the
right to the right.  Clearly $n_1+n_2+n_3+n_4=2 N$. Moreover $n_1+n_2 = n_3 + n_4$ and $n_1+n_3=n_2+n_4$, as there are equal number of particles on the two sides of the tagged particle at both the initial and final times. These equalities imply $n_1=n_4$ and $n_2=n_3$.  The number of ways of choosing the set $\{n_1,n_2,n_3,n_4\}$ is given by the multinomial coefficient  $$\frac{(2N)!}{n_1! n_2! n_3! n_4!}, $$ and each possibility occurs with probability $$ p_{--}^{n_1} p_{-+}^{n_2} p_{+-}^{n_3} p_{++}^{n_4}.$$ Hence,  summing over all possible values of $\{n_1,n_2,n_3,n_4\}$ we get 
\begin{equation}
F_{1N} =
\sum_{n_1+n_2+n_3+n_4=2N} \frac{(2N)!}{n_1! n_2! n_3! n_4!} \,
p_{--}^{n_1} p_{-+}^{n_2} p_{+-}^{n_3} p_{++}^{n_4}\,
\delta_{n_1, n_4} \delta_{n_2, n_3}\,
\end{equation}
Now, after using the integral representation of the
Kronecker delta,
\begin{equation}
\delta_{m,n}=\frac{1}{2\pi} \int_{-\pi}^{\pi} e^{i(m-n)\theta} \, d\theta 
\end{equation}
in the above equation, it immediately follows that
\begin{subequations}
\begin{align}
F_{1N}(x,y,t) &=
\int_{-\pi}^{\pi} \frac{d\phi}{2\pi}\int_{-\pi}^{\pi}\frac{d\theta}{2\pi} 
~\Bigl[p_{++}(x,y,t) e^{i\phi} + p_{--}(x,y,t) e^{-i\phi} + p_{+-}(x,y,t) e^{i\theta} + p_{-+}(x,y,t) e^{-i\theta}\Bigr]^{2N} 
\\
&= \int_{-\pi}^{\pi} \frac{d\phi}{2\pi} \int_{-\pi}^{\pi} \frac{d\theta}{2\pi}
~\Bigl[1-(1-\cos{\phi}) ~(p_{++}+ p_{--})+i \sin \phi ~(p_{++}-p_{--}) \notag \\
&\qquad\qquad\qquad\qquad\qquad\qquad
-(1-\cos {\theta})~ 
(p_{+-} + p_{-+} ) ~ +i \sin \theta~(p_{+-}-p_{-+}) \Bigr]^{2N}~ \\
&=\int_{-\pi/2}^{\pi/2} \frac{d\phi}{\pi} \int_{-\pi}^{\pi} \frac{d\phi}{2\pi} 
~\Bigl[1-(1-\cos{\phi}) ~(p_{++}+ p_{--})+i \sin \phi ~(p_{++}-p_{--}) \notag \\
&\qquad\qquad\qquad\qquad\qquad\qquad
-(1-\cos {\theta})~ 
(p_{+-} + p_{-+} ) ~ +i \sin \theta~(p_{+-}-p_{-+}) \Bigr]^{2N}~. 
\end{align}
\end{subequations}
In the last step the range of the $\phi$ integral has been broken into
two parts, and since $2N$ is even, each of these contributes equally.

\section{Evaluation of $F_{2N}(x,y,\tilde{x},\tilde{y},t)$}

Let $F_{2N}(x,y, \tilde x,\tilde y, t)$ be the probability that there
are an equal number of particles on both sides of $x$ and $y$ at $t=0$
and $t$ respectively, {\emph{given}} that there is one particle at $(\tilde x,
0)$ and another one at $(\tilde y, t)$. The particle from $\tilde{x}$
goes to $y$ and the particle from $x$ goes to $\tilde{y}$ in time
$t$. To compute $F_{2N}$, one has to keep track of both these
particles. There arises four situations (a) $\tilde x <x$ and
$\tilde{y} < y$, (b) $\tilde x >x$ and $\tilde{y} > y$, (c) $\tilde x
<x$ and $\tilde{y} > y$, and (d) $\tilde x >x$ and $\tilde{y} < y$.

Let there be $n_1$ particles going from the left of $x$ to the left of
$y$, $n_2$ particles from the left to the right, $n_3$ particles from
the right to the left, and $n_4$ particles from the right to the
right.  Since two of the particles are considered separately, the rest
can be chosen in $(2N-1)!/(n_1! n_2! n_3!  n_4!)$ different ways and
$n_1+n_2+n_3+n_4=2N-1$.

Now, in the situation (a) we have, $n_1+n_2+1=n_3+n_4$ and
$n_1+n_3+1=n_2+n_4$.  These conditions are equivalent to $n_2=n_4$ and
$n_1=n_4-1$. Similarly, the conditions for the other
three situations can be worked out, and this gives (b) $n_1=n_4$ and $n_2=n_3-1$, (c)
$n_2=n_3$ and $n_1=n_4+1$, and (d) $n_1=n_4$ and $n_2=n_3+1$,
respectively.  

Now following the procedure used to evaluate $F_{1N}$,
it is easily found that
\begin{subequations}
\begin{align}
F_{2N} (x,y,\tilde{x},\tilde{y},t) &=
\int_{-\pi}^\pi\frac{d\phi}{2\pi}\int_{-\pi}^\pi \frac{d\theta}{2\pi}
[p_{++} e^{i\phi} + p_{--} e^{-i\phi} + p_{+-} e^{i\theta} + p_{-+}
  e^{-i\theta}]^{2N-1}\, \psi(\theta,\phi),\\
&=
\int_{-\pi/2}^{\pi/2}\frac{d\phi}{\pi}\int_{-\pi}^\pi \frac{d\theta}{2\pi}
[p_{++} e^{i\phi} + p_{--} e^{-i\phi} + p_{+-} e^{i\theta} + p_{-+}
  e^{-i\theta}]^{2N-1}\, \psi(\theta,\phi),
\end{align}
\end{subequations}
where the extra phase factor $\psi(\theta,\phi)$ originates from
addend $\pm1$ that appear in the relations among $n_i$'s above, and
$\psi(\theta,\phi)=e^{-i\phi}$, $e^{i\phi}$, $e^{-i\theta}$, and
$e^{i\theta}$ for the situations (a) $\tilde x <x$ and $\tilde{y} <
y$, (b) $\tilde x >x$ and $\tilde{y} > y$, (c) $\tilde x <x$ and
$\tilde{y} > y$, and (d) $\tilde x >x$ and $\tilde{y} < y$
respectively. To arrive at the last line, we have broken the integral
over $\phi$ into two parts, and used the fact that $2N-1$ is odd, and
phase factors yield a extra factor of $(-1)$ when the phases are
shifted by $\pi$.

\end{document}